\documentclass[twocolumn,showpacs,preprintnumbers,amsmath,amssymb]{revtex4}

\usepackage{graphicx,epsfig}
\usepackage{dcolumn}
\usepackage{bm}

\begin{document}

\title{A minimal model of parallel electric field generation in a transversely inhomogeneous plasma}
\author{David Tsiklauri}
\affiliation{Institute for Materials Research,
University of Salford, Greater Manchester, M5 4WT, United Kingdom.}

\date{\today}

\begin{abstract}
We study the generation of parallel electric fields by virtue of propagation of ion cyclotron waves (with
frequency 0.3 $\omega_{ci}$) in the plasma
with a transverse density inhomogeneity. Using two-fluid, cold plasma 
linearised equations, we show for the first time 
that $E_{\parallel}$ generation 
can be understood by an analytic equation that couples $E_{\parallel}$  to the transverse 
electric field of the driving ion cyclotron wave.
We prove that the minimal model required to reproduce previous kinetic results
on $E_{\parallel}$ generation is the two-fluid, cold plasma approximation in the linear regime. In this simplified model,
the generated $E_{\parallel}$  amplitude e.g. 
for plausible solar coronal parameters attains values of  $\approx 90$ Vm$^{-1}$ for the
mass ratio $m_i/m_e=262$, within a time corresponding to 3 periods of the driving ion cyclotron wave.
By considering the  numerical solutions we also show that the cause of $E_{\parallel}$ generation is
electron and ion flow separation (which is not the same as electrostatic charge separation)
induced by the transverse density inhomogeneity. The model also correctly reproduces the previous kinetic results 
in that only electrons are accelerated (along the background magnetic field), while ions do not accelerate 
substantially. We also investigate how $E_{\parallel}$ generation is affected by the mass ratio and found
that amplitude attained by $E_{\parallel}$ decreases linearly as inverse of the mass ratio $m_i/m_e$, i.e. 
$E_{\parallel} \propto 1/m_i$. This result contradicts to the earlier suggestion by G\'enot et al (1999, 2004)
that the cause of $E_{\parallel}$ generation is the polarisation drift of the driving wave, which scales as $\propto m_i$.
Also, for realistic mass ratio of $m_i/m_e=1836$ our empirical scaling law is producing  $E_{\parallel}=14$ Vm$^{-1}$ (for solar coronal
parameters). 
Increase in mass ratio does not have any effect on final 
parallel (magnetic field aligned) speed attained by electrons. However, parallel ion
velocity decreases linearly with inverse of the mass ratio $m_i/m_e$, i.e. parallel 
velocity ratio of electrons and ions scales directly as $m_i/m_e$. 
These results can be interpreted as following: (i) ion dynamics plays no role in the $E_{\parallel}$
generation; (ii) decrease in the generated parallel electric field amplitude with the increase of the mass ratio $m_i/m_e$
is caused by the fact that $\omega_d = 0.3 \omega_{ci} \propto 1/m_i$ is decreasing, and hence the electron fluid can
effectively "short-circuit" (recombine with) the slowly 
oscillating ions, hence producing smaller $E_{\parallel}$ which also scales exactly as $1/m_i$.
\end{abstract}


\maketitle

\section{Introduction and Motivation}

The generation of parallel electric fields in inhomogeneous plasmas is a generic
topic, which is of interest in a variety of plasma phenomena such as 
particle acceleration in Solar and stellar flares \cite{lf05}, auroral acceleration
region and current sheets in the Earth magnetosphere (see refs. in \cite{sl06}), laboratory plasma reconnection 
experiments \cite{yam97,matt05} and many more.
In situ and remote observations of accelerated particles often show
parallel electric fields in localised double layers, charge holes or U-shaped voltage drops.

In many astrophysical plasmas, an adequate form of description of large-scale, bulk dynamics 
is provided by Magnetohydrodynamics (MHD). However, MHD cannot provide proper
description of some fundamental questions such as dissipation 
(which necessarily occurs at small-scales) and particle acceleration, unless the concept of
somewhat uncertain from the fundamental point of view anomalous resistivity is invoked.
The particle acceleration is of a considerable importance e.g. for Solar flares where
the accelerated particles gain 50-80\% of the energy released during this
process. On one hand, {\it observable} dynamics e.g. (i) MHD waves in the case Solar plasmas; (ii) jets and
accretion disks, in the case of stellar or compact objects or centres of Galaxies; and (iii) MHD waves in
Tokamak spectroscopic studies; are well described by MHD theory. On the other hand,
small-scale processes such as dissipation and particle acceleration are  {\it not observable}
directly. This creates controversy around issues such as the coronal heating problem (as to why
the Solar corona is 200 times hotter than underlying photosphere); anomalous resistivity which
manifests itself in an unusually fast damping of kink oscillations of solar coronal loops; or anomalous
viscosity (problem of getting rid of angular momentum) in accretion disks.
This dichotomy is schematically sketched in Fig.~1. Here energy cascade from the
large scales to small scales is depicted as either $1/f=k^{-1}$ the white noise spectrum (in the case of waves) or 
some form of turbulence spectrum (with some power law of $k^{-\alpha}$ dependent on a particular turbulence model).

When MHD is used for the description of plasmas, the electric field is totally eliminated
from the consideration. On one hand, this  has a good justification due to the condition:
\begin{equation}
\frac{1}{c}\frac{\partial \vec E}{\partial t} / \left( \nabla \times \vec B\right) \approx
\frac{1}{c}\frac{E}{T}\frac{L}{B}=\frac{1}{c}\frac{VB}{cT}\frac{L}{B}=\left(\frac{V}{c}\right)^2\ll 1,
\end{equation}
i.e. for non-relativistic $(V \ll c)$ plasmas the ratio of the displacement and $( \nabla \times \vec B)$
currents is much smaller than unity. Note that in the Eq.(1) spatial and time derivatives were
approximated by: $\partial / \partial x \approx 1/L$ and
$\partial / \partial t \approx 1/T$, where $L$ and $T$ are typical spatial and temporal
scales of the system; and the ideal MHD limit ($\vec E=-\vec V \times \vec B /c$) was used.
Thus, by neglecting the displacement current the electric field is totally excluded 
from the consideration. On one hand, this assumption 
may well be valid for the large scales. On the other hand, when the small scales are
considered, the electric field, which appears (as we will show below) because
of the electron and ion flow separation (which is impossible to treat correctly in {\it single} fluid MHD)
starts to play far more important role than previously thought.

\begin{figure}[]
\resizebox{\hsize}{!}{\includegraphics{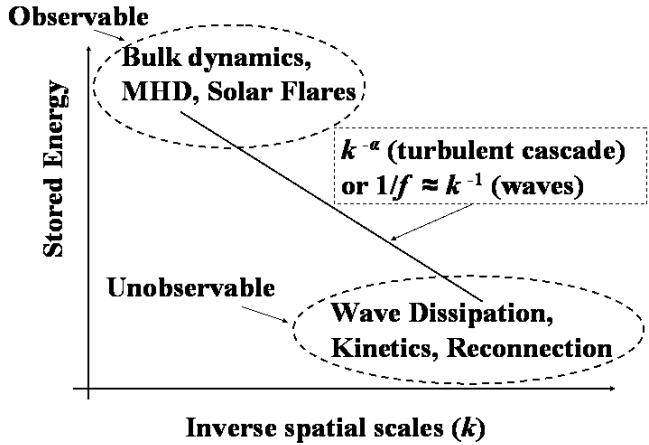}} 
\caption{The sketch of typical power law spectrum of fluctuations along with their observability criteria.}
\end{figure}

Authors of Ref.~\cite{sl06} pointed out that previous studies of the $E_{\parallel}$
generation, based on the balance of the different terms in the
generalised Ohm's law, were not properly addressing the issue.
In essence their main argument was that in such approach the
generalised Ohm's law merely states the Newton's second law $F=ma$, whilst
obscuring the true source of the parallel electric field generation.
It was suggested that the source of $E_{\parallel}$ is the parallel
displacement current. As stated above, this term is usually ignored, however
in the regions of low density, for a certain $( \nabla \times \vec B)_\parallel$,
the plasma is too dilute to carry significant $J_\parallel$ and thus 
$(1/c) \partial E_\parallel / \partial t$ becomes important \cite{sl06}.
One of the main conclusions that immediately follows is that the signatures of
the generated $E_{\parallel}$ in space plasmas should be correlated with
low plasma density.

Yet another series of
works exist, which investigate the generation of parallel electric fields by
virtue of propagation of Alfv\'en waves (or more precisely ion-cyclotron waves, see below) in the plasma
with a transverse density inhomogeneity \cite{glq99,glm04,tss05a,tss05b,mgl06,t06a,t06b}.
To this day the true cause of the generation of $E_{\parallel}$ in these
studies eluded determination. Authors of Ref.~\cite{glq99} considered the case
of both transverse and longitudinal density inhomogeneity, applicable to
the stratified Earth magnetosphere. They demonstrated that
$E_{\parallel}$ is generated in the regions of transverse density
gradients, and presented an analytical model in which 
the $E_{\parallel}$ and $E_{\perp}$  are coupled via 
{\it longitudinal} density gradient (see Eq.(6) from Ref.~\cite{glq99}).
Subsequently, detailed numerical study of long term evolution of the system 
was presented, including the generation of $E_{\parallel}$ \cite{glm04}.
However, in Ref.\cite{glm04} only the case of transverse density inhomogeneity was
considered, while theoretical explanation was still based on 
Ref.~\cite{glq99}. This seems incorrect because the latter reference
attributes $E_{\parallel}$ and $E_{\perp}$ coupling to the longitudinal
inhomogeneity, which is absent in Ref.~\cite{glm04}. In brief, these two works
suggest that the Alfv\'en wave propagation
on sharp density gradients leads to the formation of a significant
parallel electric field.
It results from an electric charge separation generated on the
density gradients by the polarisation drift associated with the
time varying Alfv\'en wave electric field \cite{glm04}. Their approach involved
substituting ion polarisation drift current (electron one was omitted 
because of its proportionality to the particle mass)
$j_\perp=(m_i n_i /B^2) \partial E_\perp / \partial t$
into the Maxwell equations, which with the aid of the
conservation laws yielded the equation for 
$E_{\parallel}$ and  $E_{\perp}$ coupling \cite{glq99}.
Unaware of these works authors of Refs.~\cite{tss05a,tss05b}
considered similar physical system with the increased
density in the middle of the domain (mimicking) solar coronal
loop, as opposed to Earth magnetospheric density cavity case
studied in Refs.~\cite{glq99,glm04}. Similar effect of $E_{\parallel}$
generation was found because of the existence of density gradients in the
system. Later a comment paper was published \cite{mgl06}, which
detailed similarities and differences of the two series of works.

It should be noted in passing that at that time we came to the
realisation that electron acceleration seen in both series of
works \cite{glq99,glm04,tss05a,tss05b} is a non-resonant
wave-particle interaction effect. In Refs.~\cite{tss05a,tss05b} 
the electron thermal speed was $v_{th,e} = 0.1c$ while
the Alfv\'en speed in the strongest density gradient regions was
$v_A = 0.16c$; this unfortunate coincidence led us to the conclusion
that the electron acceleration by parallel electric fields
was affected by the Landau resonance with the phase-mixed
Alfv\'en wave. In Refs.~\cite{glq99,glm04} the electron
thermal speed was $v_{th,e} = 0.1c$ while the Alfv\'en speed
was $v_A = 0.4c$ because they considered a more strongly magnetised
plasma applicable to Earth magnetospheric conditions.
Based on this observation, Refs.~\cite{t06a,t06b} explored the possibility
of $E_{\parallel}$ generation in the MHD description in the solar
coronal heating problem context. Although, in the latter approach, 
the heating aspect seems
certain (because the fast magnetosonic waves, which are generated by the
interaction of weakly non-linear Alfv\'en wave with the 
transverse density inhomogeneity, dissipate on the bulk Braginkii resistivity),
the issue whether such $E_{\parallel}$ can accelerate particles is
less clear \cite{t06b}.

\section{The model and results}

The above discussion demonstrates that the issue of true cause of
$E_{\parallel}$ generation when an Alfv\'en wave moves in the
transversely inhomogeneous plasma eluded identification.
In this work we present a minimal model which can explain 
$E_{\parallel}$ generation in mathematically and physically 
rigorous manner. We start from two-fluid, cold (ignoring thermal
pressure) plasma linearised equations \cite{kt73}:
\begin{equation}
\partial_t \vec V_e=-(e/m_e)\left(\vec E + \vec V_e \times \vec B_0 /c \right),\\
\end{equation}
\begin{equation}
\partial_t \vec V_i=+(e/m_i)\left(\vec E + \vec V_i \times \vec B_0 /c \right),
\end{equation}
\begin{equation}
\partial_t \vec B= -c \nabla \times \vec E,
\end{equation}
\begin{equation}
\partial_t \vec E = c \nabla \times \vec B - 4 \pi n e (\vec V_i - \vec V_e).
\end{equation}
Hereafter subscripts under $\partial$ denote partial derivative with respect to that
subscript. Uniform, background magnetic field, $B_0$ is in $z$-direction.
Density profile is specified as a ramp, 
$n(x)=n_0\left(1+3 \exp \left[-[(x-100\delta)/ (20 \delta)]^6 \right]\right)$
in which the central region (along $x$-direction, i.e. across $z$),
is smoothly enhanced by a factor of 4, and there
are the strongest density gradients having a width of about $20 \delta$
around the points $x = 81 \delta$ and $x = 119\delta$.
Here $\delta=c/\omega_{pe}$ is the (electron) skin depth, which is a
unit of  grid in our numerical simulation.
We use 2.5D description meaning that we keep all three, $x,y,z$ components
of all vectors, however spatial derivatives $\partial / \partial y \equiv 0$.
The above normalised plasma number density and Alfv\'en speed profiles are shown in Fig.(2).

In order to derive  the equation that describes  $E_{\parallel}=E_z$
generation, we write Eqs.(2)-(5) in $x,y,z$ component form. Omitting details of the
calculation we present the final result:
\begin{equation}
\left(\partial^2_{tt} - c^2 \partial_{xx}^2 + \omega_{pi}^2+ \omega_{pe}^2 \right) E_{\parallel}=
-c^2 \partial^2_{zx} E_x.
\end{equation}
Also, a similar calculation enables us to obtain the equation describing the dynamics 
of driving transverse electric field $E_x$ of an ion cyclotron wave:
$$
\left(\partial^2_{tt} - c^2 \partial_{zz}^2 + \omega_{pi}^2+ \omega_{pe}^2 \right) E_x=
$$
\begin{equation}
-c^2 \partial^2_{zx} E_{\parallel}-
\omega_{pi}^2(m_i/e)\omega_{ci}V_{iy}-\omega_{pe}^2(m_e/e)\omega_{ce}V_{ey}.
\end{equation}
In the considered problem $E_x$ and $B_y$ are both components of Alfv\'en (ion cyclotron) wave, so these can be used
interchangeably.
Note that Eq.(7) also describes the feedback of the generated $E_{\parallel}$ on the
driving transverse electric field $E_x$ (see the first term on the right-hand-side).
Here the notation is standard: $\omega_{pe}=\sqrt{4 \pi n_0 e^2/m_e}$ and $\omega_{pi}=\sqrt{4 \pi n_0 e^2/m_i}$  are
electron and ion plasma frequencies;  $\omega_{c(e,i)}=eB_0/(m_{(e,i)}c)$ are respective cyclotron frequencies.

It is interesting to note that Eqs.(6) and (7) can be also obtained from the dielectric
permeability tensor of cold, magnetised plasma (e.g. chapter 4.9 in Ref.\cite{kt73}).
For example, Eq.(6) can be directly obtained from the classical equation for the electric field perturbation, $\vec E_1$, in the case of
$\vec E_0=0$ and $\vec B_0=B_0 \hat z$
\begin{equation}
\nabla \times \nabla \times \vec E_1 =(\omega^2 / c^2) \breve \varepsilon  \vec E_1,
\end{equation}
where $\breve \varepsilon$ is the dielectric
permeability tensor of cold, magnetised plasma. In effect, Eq.(6) can be obtained from the $z$-component of Eq.(8) and
putting in $\varepsilon_{zz}=1-\omega_{pe}^2 / \omega^2 -\omega_{pi}^2 / \omega^2$.

\begin{figure}[]
\resizebox{\hsize}{!}{\includegraphics{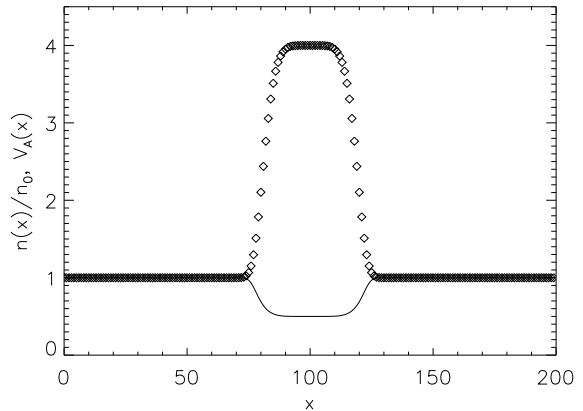}} 
\caption{Dimensionless number density, open squares, and Alfv\'en speed, solid line, profiles across the uniform unperturbed magnetic field 
(i.e. along $x$-coordinate) which is
used as an equilibrium configuration in our model of a footpoint of a solar coronal loop or a polar region plume.}
\end{figure}

In order to solve Eqs.(2)-(5) numerically we use the following normalisation:
$t=\tilde t \omega_{pe}^{-1}$, $V_{x,y,z}=\tilde V_{x,y,z} c$, $E_{x,y,z}=\tilde E_{x,y,z} (m_e c \omega_{pe} / e) =
\tilde E_{x,y,z}  E_0$,
$B_{x,y,z}=\tilde B_{x,y,z} B_0$, and $(x,y,z)= c/\omega_{pe} (\tilde x, \tilde y, \tilde z)$.
In what follows we omit tilde on the dimensionless quantities.
The $(x,z)$ simulation 2D box size is $200 \delta \times 2500 \delta$. Since we fix background 
plasma number density at $10^9$ cm$^{-3}$ (typical value for the solar corona), 
$\omega_{pe}$ is then $1.784 \times 10^9$ rad s$^{-1}$ and the simulation box size is 
$33.6$ m in $x$-  and $420.5$ m in $z$-direction. $B_0$ was fixed at 101.5 Gauss 
(typical value for the solar corona), which gives $\omega_{ce} / \omega_{pe} = 1$.
$m_i / m_e$ ratio was varied as: 45.9, 91.8, 183.6, 262.286 (realistic one is 1836).
These values correspond to $1/40,\; 1/20, \; 1/10$ and $1/7$-th of the realistic value
respectively.
This yields respectively: 
$\omega_{ci} / \omega_{pi} = B_0/(c \sqrt{4 \pi n_i m_i})=V_A / c=1/\sqrt{m_i/m_e}=0.148, \;0.104,\;0.074$
and $0.062$ for $x\leq 70$
and  $x\geq 130$ (realistic $\omega_{ci} / \omega_{pi} =V_A / c$ is 0.023).
Here parameters are similar to e.g. Refs.\cite{tss05a,tss05b}, except for
far more realistic mass ratios. Note that the simulation parameters are still
somewhat artificial. 
Full kinetic, Particle-In-Cell (PIC) simulations employed in Refs.\cite{tss05a,tss05b} or
in gyro-kinetic approach which uses guiding centre approximation  for electrons, whilst retaining ion particle-like dynamics \cite{glq99,glm04}
are computationally challenging. 
Thus, in those studies rather modest mass ratios e.g. 16 were used.
Note also, that since here we do not need to resolve electron 
thermal motions as we are only studying electromagnetic part of the 
problem ($E_{\parallel}$ generation) our unit of spatial grid size is $\delta=c/\omega_{pe}$, the (electron) skin depth.
While in full kinetic, PIC simulation \cite{tss05a,tss05b} the unit of grid has to be $\Delta=v_{th,e}/\omega_{pe}$.
Since in a PIC simulation typically $v_{th,e} / c =0.1$, in the present, two-fluid approach an equivalent to PIC numerical simulation 
requires $(\delta/ \Delta)^2=(c/v_{th,e})^2=10^2$ less grid points, thus it can be 100 times faster. 
For comparison a single run for mass ration 16 
in Refs.\cite{tss05a,tss05b} takes about 8 days on parallel, 32 dual-core 2.4 GHz Xeon processors,
similar run with mass ratio of 262 would have taken 4 months.
The numerical run presented here for the mass ratio of 262 takes 4 days with only one processor.

We solve {\it relativistic} version of
Eqs.(2)-(5) numerically with a specially developed and tested FORTRAN 90 code
which uses 4-th order centred spatial derivatives and 4-th order Runge-Kutta time marching. 
Although Alfv\'en speeds considered are at most $\approx 15$ \% of the speed of light for $m_i/m_e=45.9$,
relativistic effects were included. The simplest option becomes available in the linear
regime. In ref.  \cite{kt73}, appendix I, paragraph 5, it was shown that the
relativistic equation of motion of a particle with charge $q$ and the rest mass $m_0$ 
can be written as
\begin{equation}
\frac{d}{dt}\vec V =\frac{q}{\gamma m_0} \left[\vec E +\frac{\vec V \times \vec B}{c} - 
\frac{\vec V (\vec V \cdot \vec E)}{c^2} \right],
\end{equation}
where $\gamma=(1-V^2/c^2)^{-1/2}$. As can be seen from the latter equation, in the linear regime, it
coincides with either Eq.(2) or (3) after substituting $m_{e,i} \to \gamma_{e,i} m_{e,i}$, where
$\gamma_{e,i}=(1-V_{e,i}^2/c^2)^{-1/2}$. Naturally, such simplified approach is only valid 
when there are no flows in the unperturbed state $V_0=0$. As can be seen below, largest attained velocities
in the simulation are those of electrons, and these do not exceed 3 \% of speed of light. Thus,
relativistic corrections play only a minor role. It should be noted, however we still retain the displacement current
in Eq.(5).
Note, also that the gradients in the code are resolved numerically to an 
appropriate precision (20 grid points (open squares) across each gradient in  Fig.(2)).

\begin{figure*}
\centering
\epsfig{file=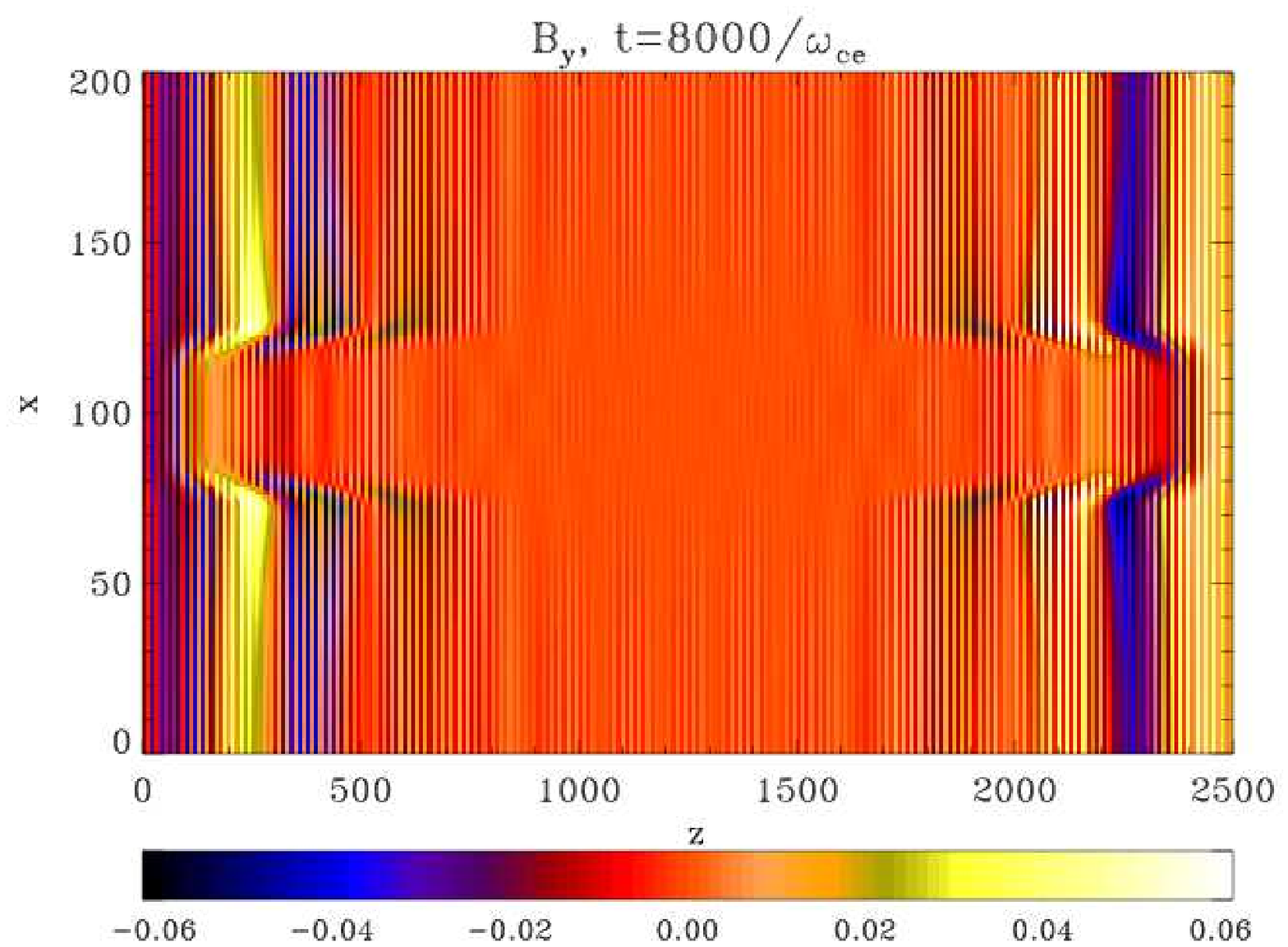,width=8cm}
\epsfig{file=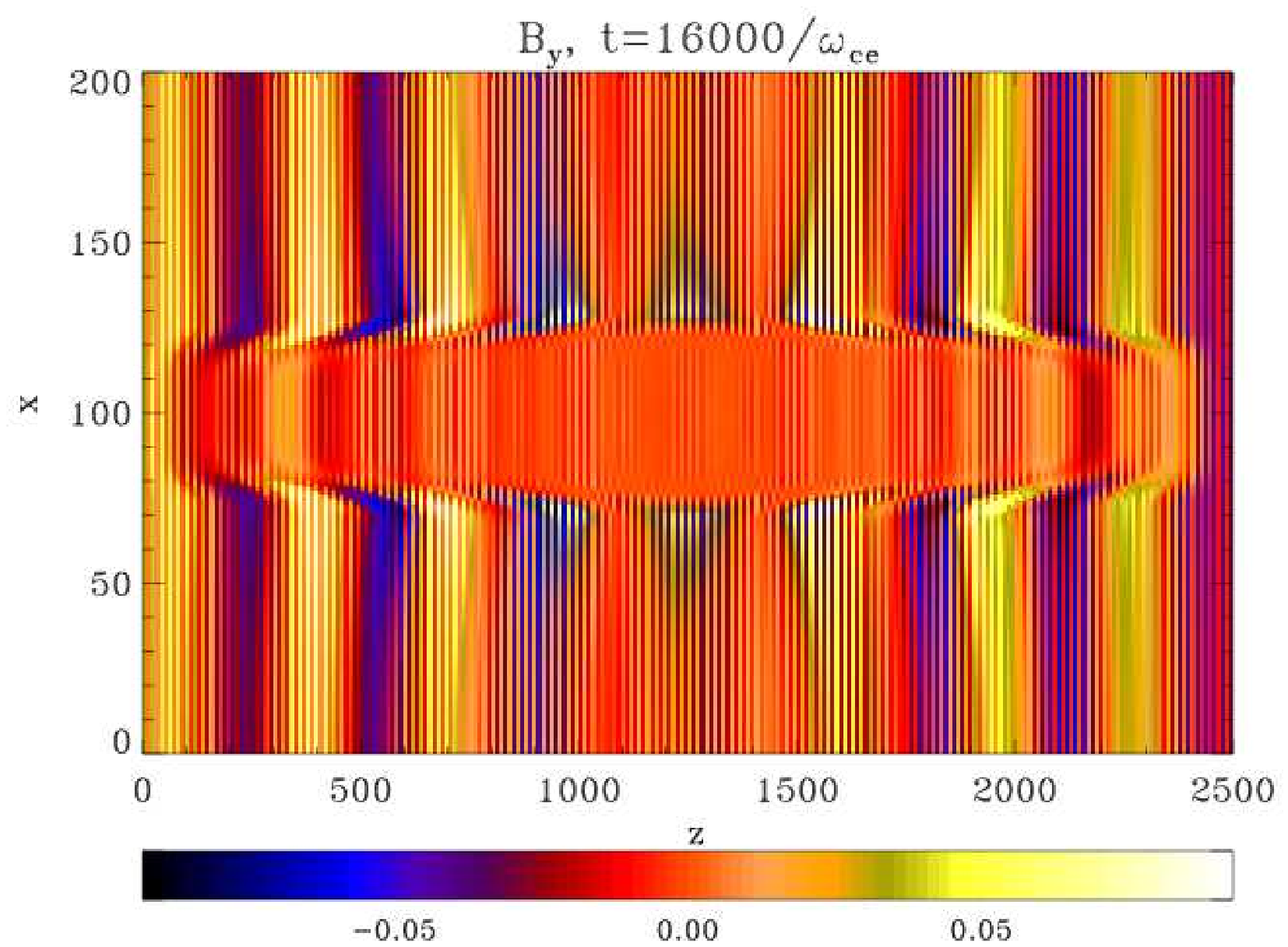,width=8cm}
 \caption{Contour (intensity) plots of  phase-mixed transverse magnetic field $B_y$ at times $t=8000 / \omega_{ce}$ (left) and 
 $t=16000 / \omega_{ce}$ (right).}
\end{figure*}

\begin{figure*}
\centering
\epsfig{file=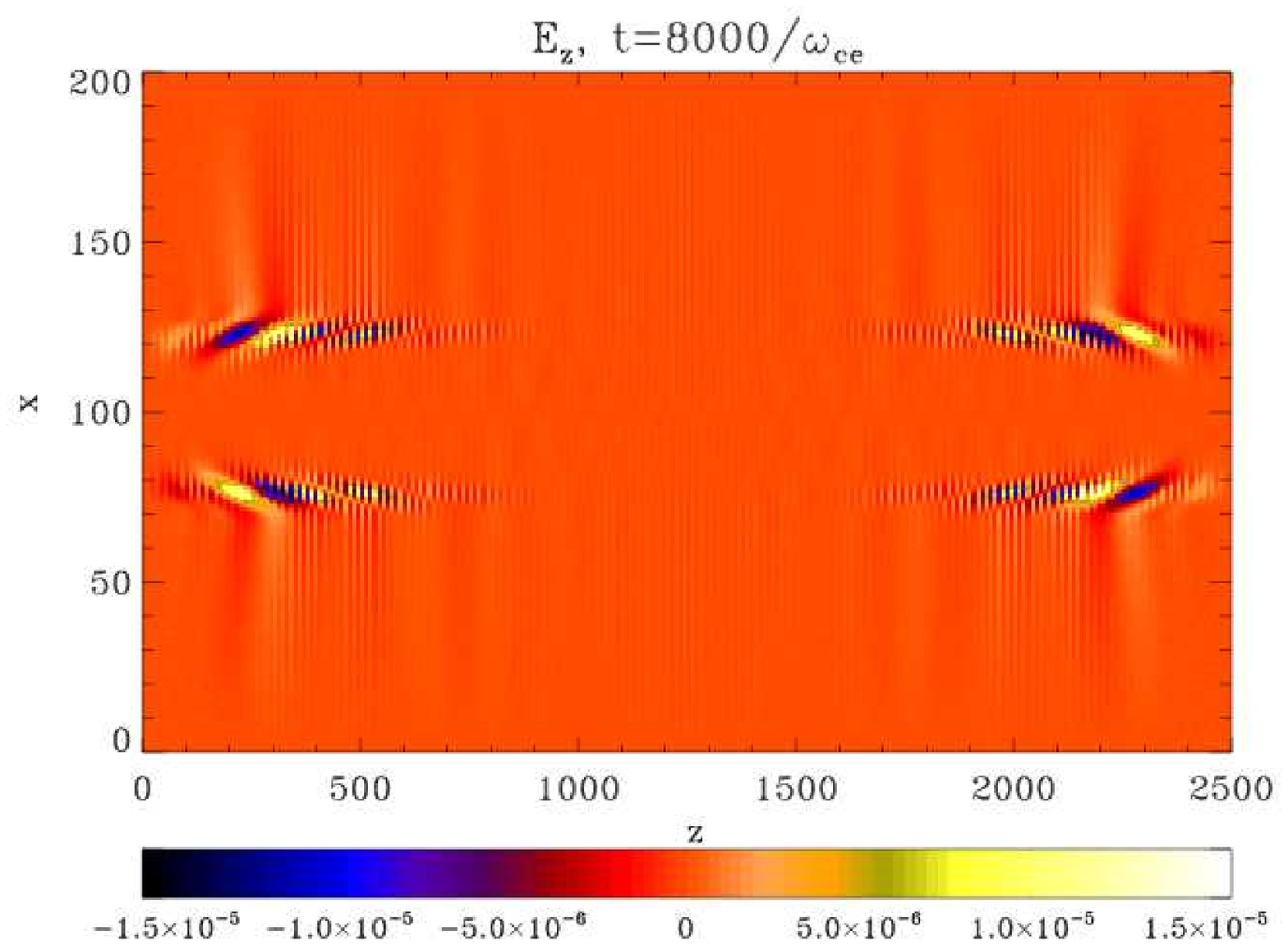,width=8cm}
\epsfig{file=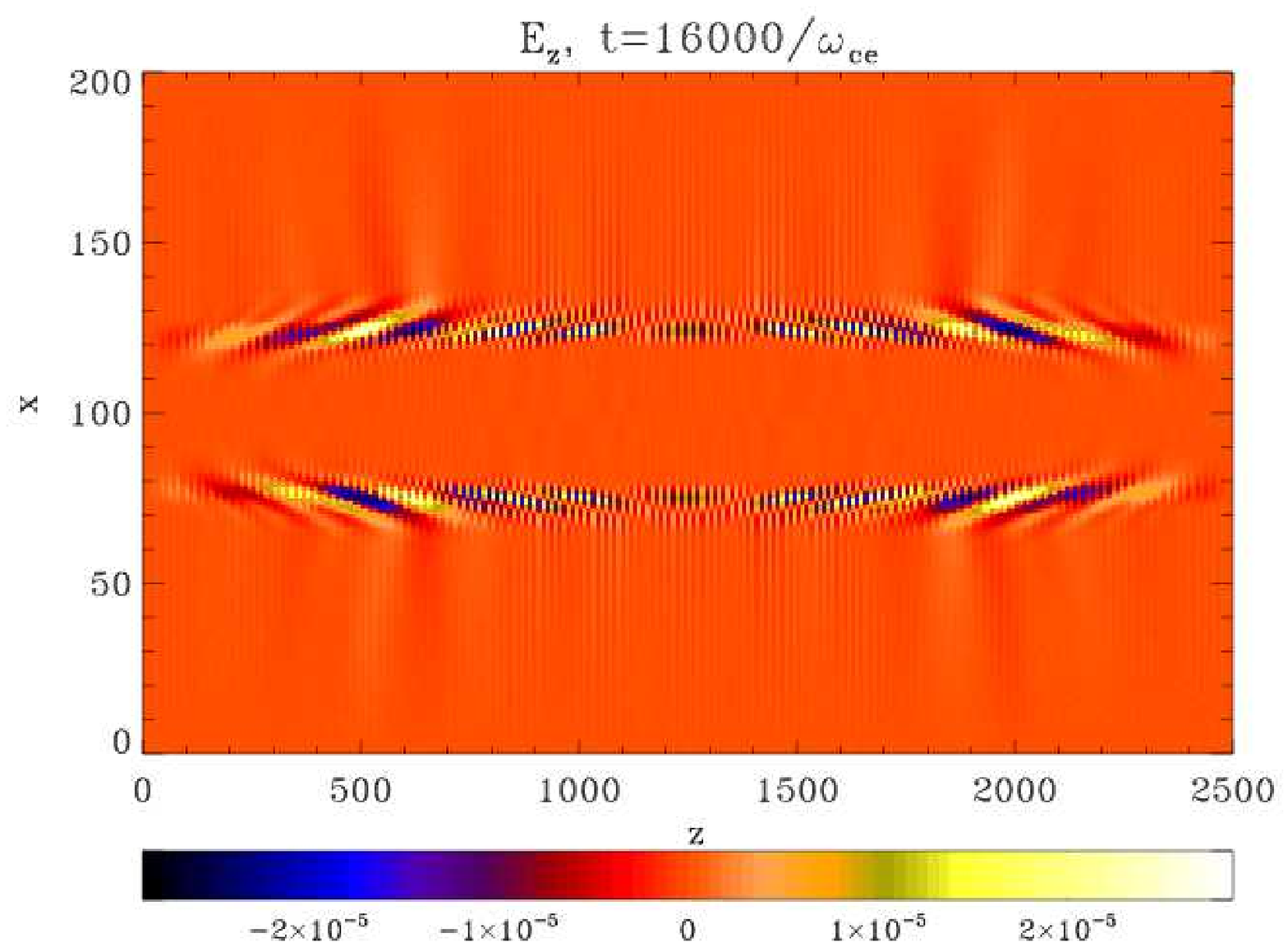,width=8cm}
\caption{Contour (intensity) plots of $E_z=E_{\parallel}$ at times $t=8000 / \omega_{ce}$ (left) and 
 $t=16000 / \omega_{ce}$ (right).}
\end{figure*}

Initially all perturbations are set to zero, and we start driving the $z=1$ cell with the transverse magnetic fields of the form
$B_y  =  -0.05 \sin(\omega_d t) \left(1.0-\exp[-(t/(3.125 \omega^{-1}_{ci}))^2]  \right)$
and $B_x =  -0.05 \cos(\omega_d t) \left(1.0-\exp[-(t/(3.125 \omega^{-1}_{ci}))^2]  \right)$. 
As in Refs.\cite{tss05a,tss05b}, we fixed $\omega_d$ at 0.3 $\omega_{ci}$ (to avoid ion-cyclotron damping playing any role).
$\left(1.0-\exp[-(t/(3.125 \omega^{-1}_{ci}))^2]  \right)$ factor ensures that these driving $B_\perp$ fields ramp up to their maximal values
in time $t=3.125 \omega_{ci}^{-1}$.
Such driving with $B_\perp$ of 5\% of the background $B_0$
excites circularly polarised ion-cyclotron (IC) waves, 
these waves are often misquoted as Alfv\'en waves \cite{glq99,glm04,tss05a,tss05b}.
For parallel to $\vec B_0$ propagation, and assuming plasma quasi-neutrality,
the relevant dispersion relation reads as  \cite{kt73}:
\begin{equation}
k^2_{R,L}=\frac{\omega^2}{c^2}\left[1-\frac{ \omega^2_{pe} + \omega^2_{pi} }{(\omega \mp \omega_{ce} )
(\omega \pm \omega_{ci} )} \right],
\end{equation}
where subscripts $R,L$ refer to the waves with right and left circular polarisation, and so are the upper and lower signs in the plus and minus.
In the frequency range $\omega \ll \omega_{ci}$ both left and right polarised waves tend to an Alfv\'en wave branch, which satisfies the following 
dispersion relation \cite{kt73}:
\begin{equation}
k^2_{R,L}=\frac{\omega^2}{c^2}\left[1+\frac{  4\pi n (m_e+m_i)c^2}{B^2} \right],
\end{equation}
which is the same as 
\begin{equation}
\frac{\omega}{k}=V_A/\sqrt{1+V^2_A/c^2}.
\end{equation}
Note, that the root in the denominator appears only because of the displacement current is kept.
At frequencies $\omega \simeq 0.3 \omega_{ci}$ however, the correct term ion-cyclotron wave instead of Alfv\'en wave 
should be used.

Note that in differ to Refs.\cite{tss05a,tss05b} we now directly drive transverse magnetic fields ($B_x,B_y$) (which in turn are coupled to transverse
electric fields ($E_x,E_y$)). Conventionally, Alfv\'enic and IC waves are more associated with magnetic field oscillation. 
However, in the kinetic, Particle-In-Cell
simulation usually variation of electric field is used for driving waves, because particles respond to electric, rather than magnetic fields 
(both are naturally coupled, such as $E_x$ and $B_y$ represent single Alfv\'en wave).
In the two-fluid numerical code like the one used here, it is possible to use 
transverse magnetic fields for driving of IC wave.

\subsection{case of $m_i/m_e = 262$}

In this subsection we consider case of $m_i/m_e=262.268 \approx 262$, which is the largest value considered
in this study.
As can be seen in Fig.(3), the generated at the left edge ($z=1$) IC waves propagate both in the directions of positive and negative $z$'s.
However, because of the periodic boundary conditions used (applied on all physical quantities) IC wave that travels to the direction of negative $z$'s (to the left)
re-appears on the right edge of the figure. As in all previous phase-mixing simulations Alfv\'en velocity is a function
of the transverse (to the background magnetic field) coordinate, $x$, i.e. $V_A=V_A(x)\propto 1/\sqrt{n(x)}$ (see Fig.(2)). Thus as can be seen in Fig.(3) the 
IC wave middle portion travels slower than the parts close
to the simulation box edge. This creates progressively strong transverse gradients and hence smaller spatial scales.
If resistive effects are included (these are absent here), such configuration usually produces greatly enhanced dissipation and IC wave amplitude decays in space as 
$\propto \exp(-z^3)$ \cite{tss05a,tss05b}. The $E_{\parallel}=E_z$ field dynamics is shown in Fig.(4). 
\begin{figure*}
\centering
\epsfig{file=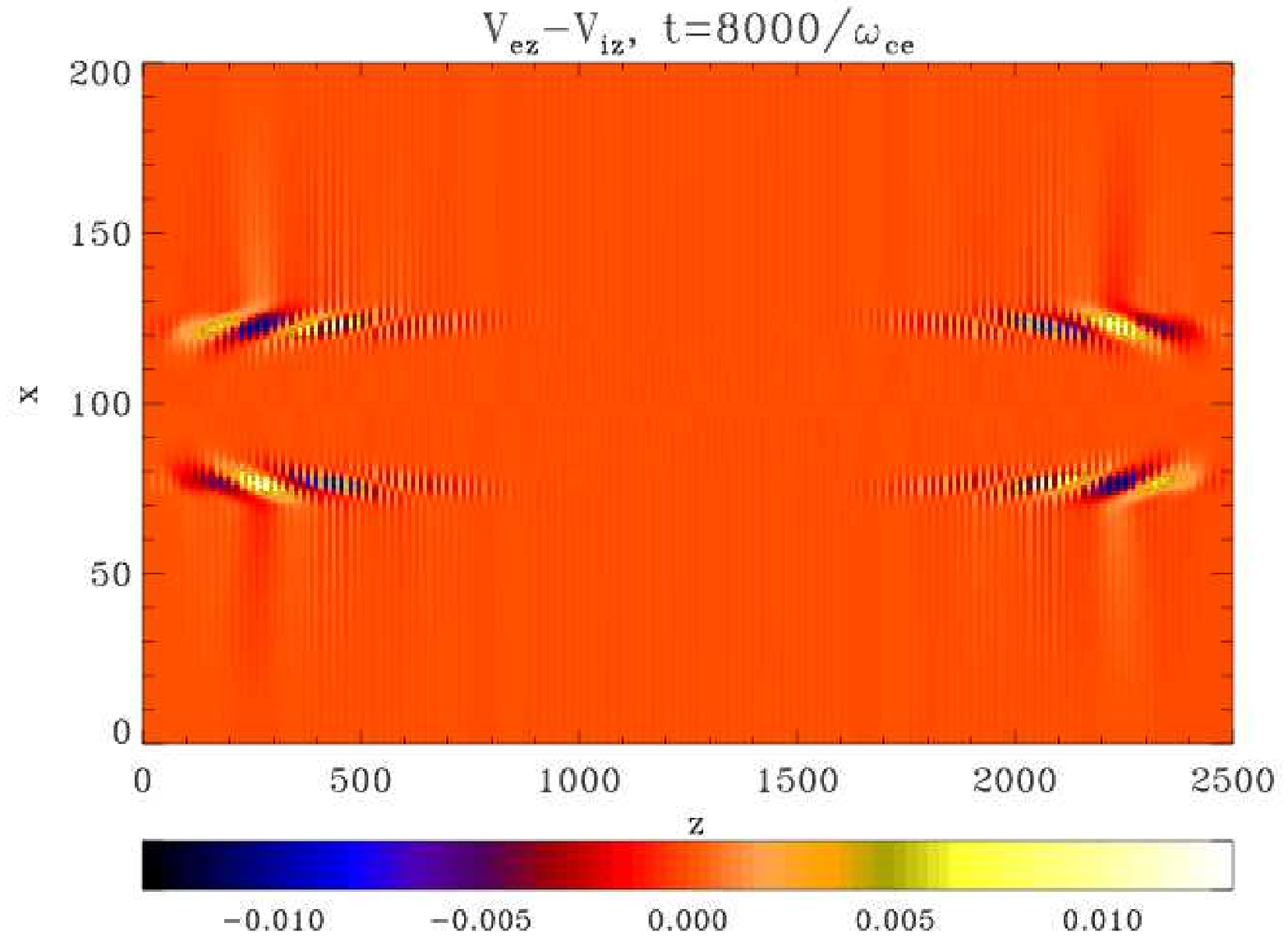,width=8cm}
\epsfig{file=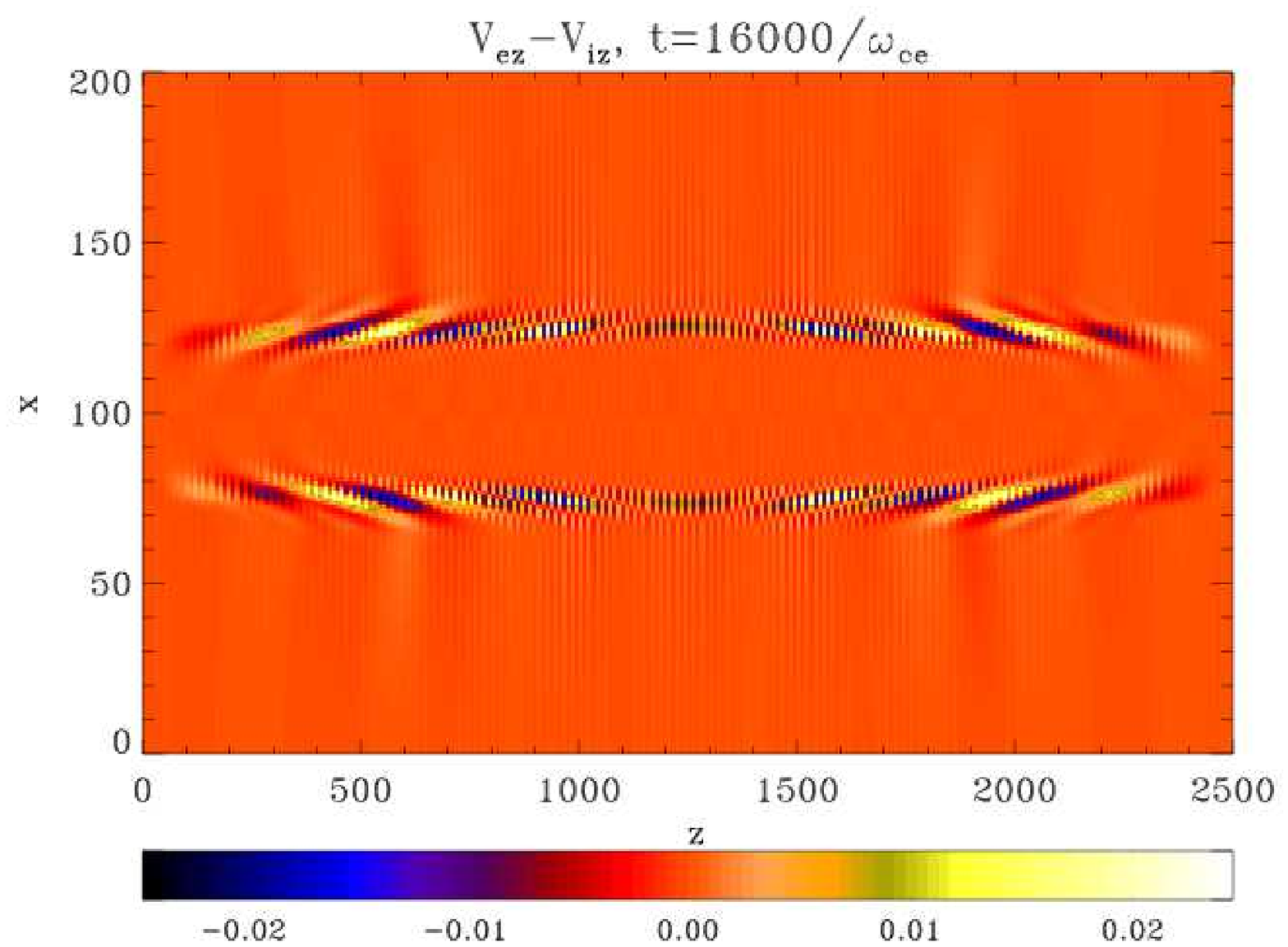,width=8cm}
 \caption{Contour (intensity) plots of $(V_{ez}-V_{iz}) \propto j_z$ at times $t=8000 / \omega_{ce}$ (left) and 
 $t=16000 / \omega_{ce}$ (right).}
\end{figure*}
We gather that  $E_{\parallel}$
is generated only in the regions of density gradients i.e. along $x = 81$ and $x = 119$ lines. This can be explained by analysing right-hand-side (RHS)
of Eq.(6). $E_{\parallel}=0$  at $t=0$ everywhere, however it can only be generated in the regions where $\partial_x E_x \not = 0$. The latter is true
only in the density gradient regions where $E_x$ becomes progressively oblique propagating.
Thus, Eq.(6), derived here for the first time, correctly explains the simulated process of $E_{\parallel}$ generation by IC waves.
It should be also noted that this equation contains $\partial_{xx}^2$, which correctly accounts for the transverse (along $x$) propagation of the
generated $E_{\parallel}$. $E_{\parallel}$'s longitudinal (along $z$) propagation due to the motion of IC wave along $z$-axis is indeed corroborated both by
Fig.(3) and Eq.(7) - note $\partial_{zz}^2$ term. 
Also, note that $E_{\parallel}$ amplitude at time $t=16000 \omega_{ce}^{-1}$ attains value of $3\times10^{-5}$.
This is somewhat smaller value than the one obtained in the full kinetic (PIC) simulation \cite{tss05a,tss05b}. 
This is due to the different mass ratios: in the the kinetic (PIC) simulation $m_i/m_e=16$, but here it is 262.
In dimensional units this $E_{\parallel}$
corresponds to about 0.003 statvolt cm$^{-1}$ or 90 V m$^{-1}$, i.e. in such electric field electrons would be accelerated to the
energy of $\approx 10$ keV over the distance of 100 m. Note, however, that the generated $E_{\parallel}$ is 
oscillatory in space and time. 

In  Fig.(5) we present $(V_{ez}-V_{iz})$ which is proportional to $j_z$, the parallel (electron and ion {\it flow separation}) current.
It can be seen from this figure that $(V_{ez}-V_{iz})$ attains moderate values of $\approx 0.03c$.
Authors of Ref.\cite{glm04} stated the importance of {\it charge separation} before. However 
the cause of $E_{\parallel}$ generation is actually electron and ion flow separation (see below). The latter is quite different from
the electrostatic effect of charge separation, which is inherently a plasma kinetic effect. Electron and ion flow separation
is a fluid-like (non-kinetic) effect, because their distribution functions remain Maxwellian at all times.

\subsection{parametric study for different $m_i/m_e$'s}

In this subsection we perform parametric study with different mass ratios $m_i/m_e$.
This is particularly useful for understanding the physics of parallel electric field
generation. This is needed because performing realistic mass ratio 
numerical simulation, $m_i/m_e=1836$, is computationally challenging.

Numerically challenging issues arise from the increase in mass ratio are the following:

(i) because our intention was to use a driving IC wave with $\omega_d=0.3 \omega_{ci}$, when changing mass ratio,
$w_d \propto 1/m_i$. Thus increasing mass of the ion leads to decrease in driving frequency.
Since in all our numerical simulations we intended to have final simulation time of 3 driving IC
wave periods, i.e.  $t_{final} = 3 \times (2 \pi / \omega_d) \propto m_i$. Thus, in turn, increase in ion mass leads to increase
in final simulation time. E.g. for the cases $m_i/m_e=$ 45.9, 91.8, 183.6, 262.286 considered here, the final simulation times were
2884, 5768, 11536 and 16000 $\omega^{-1}_{pe}$ respectively,
i.e. $t_{final} \propto m_i$.

(ii) Increase in simulation time leads to the compulsory increase in the simulation domain size, $L$, in the direction along the 
magnetic field (direction of IC wave travel), i.e.
$L = 2\times (V_A/c) \times (3 t_{final}) \propto (1/ \sqrt{m_i}) \times m_i \propto \sqrt{m_i}$

Thus, as far as the slowdown of the numerical code is concerned, the combined slowdown effect of the above two factors is $m_i\sqrt{m_i}$.

In Fig.(6) we present time evolution of the amplitude of the generated parallel electric field, which we define as 
$\max(|E_z(x,z,t)|)$, i.e. at every time step we choose one point over whole simulation domain at which modulus of 
parallel electric field is the largest. This is, as it was shown above, along the strongest gradient lines $x = 81 \delta$ and $x = 119\delta$. 
It can be seen from the graph that: (i) level attained by parallel electric field amplitude decreases with the increase in
the mass ratio; (ii) rate at which the final amplitude is reached (the averaged slope, essentially)  also 
decreases with the increase in the mass ratio.
\begin{figure}[]
\resizebox{\hsize}{!}{\includegraphics{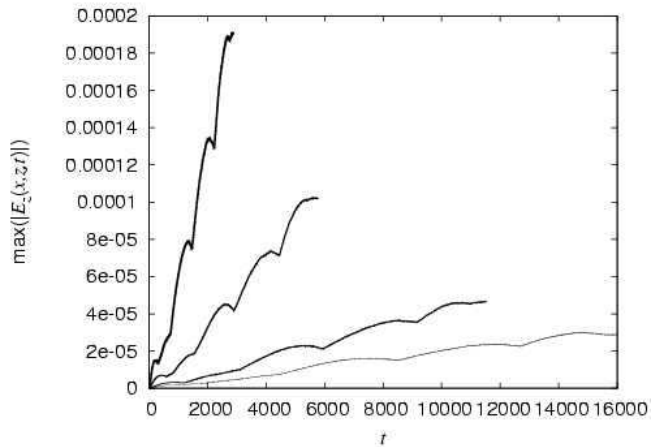}} 
\caption{Time evolution of the amplitude of the generated parallel electric field, defined as 
$\max(|E_z(x,z,t)|)$. The thickest (upper left) line corresponds to $m_i/m_e=$ 45.9; while lines with decreasing thickness correspond to 
91.8, 183.6, 262.286 respectively.}
\end{figure}

Exact behaviour of the final attained parallel electric field amplitude (within 3 periods of the driving ion cyclotron wave)
as a function of mass ratio is shown in Fig.~(7). Essentially this is a plot of the last data points (which are four) 
in the previous Fig.~(6)
as a function of $m_i/m_e$, i.e. $E^*=\max(|E_z(x,z,t_{final})|)$ vs. $m_i/m_e$. 
The dashed line corresponds to the fit $0.0085 / (m_i/m_e)$.
Plotting such graph is very useful to establish the trend. Interestingly we see that 
that amplitude attained by $E_{\parallel}$ decreases linearly with inverse of the mass ratio $m_i/m_e$.
The $x$-range in Fig.(7) is $m_i/m_e=30 - 1836$, so that rightmost point of the dashed line enables us to
grasp $E_{\parallel}$ for the case of realistic mass ratio (i.e. 1836). We thus gather that 
$E_{\parallel}=0.0085/1836= 4.630 \times 10^{-6}$ which  is $4.630 \times 10^{-6} \times E_0 = 4.7\times 10^{-4}$ statvolt cm$^{-1}$  or 
14 Vm$^{-1}$.

\begin{figure}[]
\resizebox{\hsize}{!}{\includegraphics{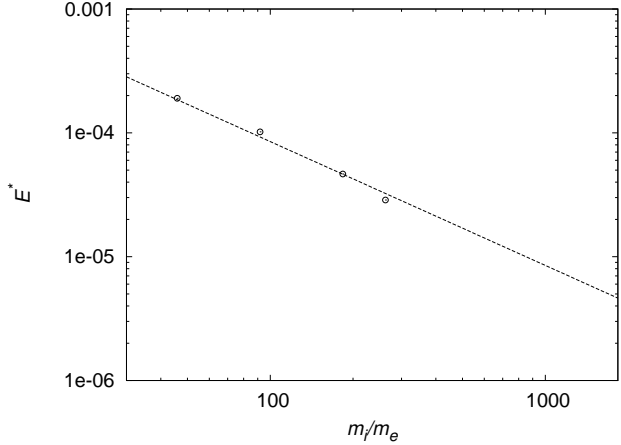}} 
\caption{The final attained parallel electric field amplitude (generated within 3 periods of the driving ion cyclotron wave)
as a function of mass ratio. Data points correspond to the last 4 data points  
in the previous Fig.~(6).
The dashed line corresponds to the fit $0.0085 / (m_i/m_e)$.
This is a log-log plot.
}
\end{figure}

In Figs.~(8) and (9) we plot the amplitudes of the generated parallel flows of electron and ion fluids which we define as 
$\max(|V_{ez}(x,z,t)|)$ and $\max(|V_{iz}(x,z,t)|)$ respectively, i.e. at every time step we choose one point over whole simulation domain at which moduli of 
parallel flows of electrons and ions are the largest. Again, this occurs somewhere along the strongest gradient lines $x = 81 \delta$ and $x = 119\delta$,
because parallel electron and ion flows are confined to the strongest density gradient regions. Four different lies in each
figure show the cases $m_i/m_e=$ 45.9, 91.8, 183.6, 262.286 considered. We gather from Figs.~(8) and (9) that
an increase in mass ratio does not have any effect on final 
parallel (magnetic field aligned) speed attained by electrons. However, parallel ion
velocity decreases linearly with inverse of the mass ratio $m_i/m_e$. 
To investigate this more quantitatively, in Fig.~(10) we plot
the ratio of final attained electron and ion flow amplitudes (within 3 periods of the driving ion cyclotron wave)
as a function of mass ratio. Essentially this is a plot of the ratio of last data points (which are four) 
in the Figs.~(8) and (9) as a function of $m_i/m_e$, i.e. $V_{ratio}=  \max(|V_{ez}(x,z,t_{final})|) /  \max(|V_{iz}(x,z,t_{final})|)$ vs. $m_i/m_e$. 
The dashed line corresponds to the fit which is $(m_i/m_e)$ with a slope of 1.
i.e. parallel 
velocity ratio of electrons and ions scales directly as $m_i/m_e$.  

\begin{figure}[]
\resizebox{\hsize}{!}{\includegraphics{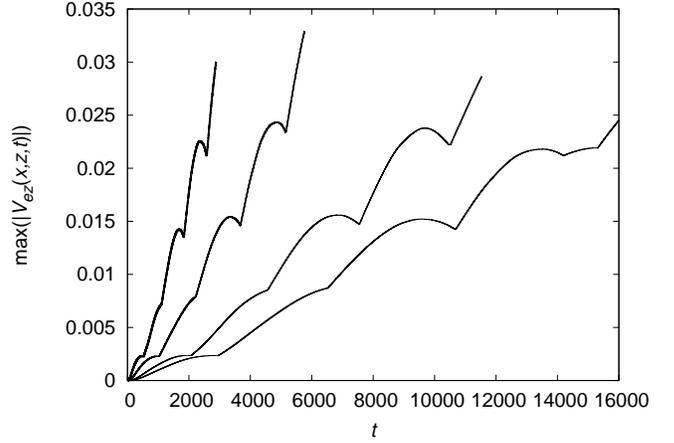}} 
\caption{Time evolution of the amplitudes of the generated parallel flows of electron fluid, defined as 
$\max(|V_{ez}(x,z,t)|)$. The thickest (upper left) line corresponds to $m_i/m_e=$ 45.9; while lines with decreasing thickness correspond to 
91.8, 183.6, 262.286 respectively.}
\end{figure}

\begin{figure}[]
\resizebox{\hsize}{!}{\includegraphics{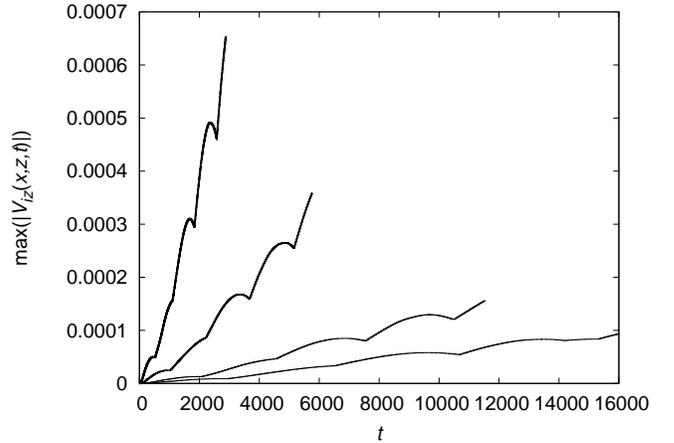}} 
\caption{The same as in Fig.~(8), but for ions.}
\end{figure}

\section{discussion}

The conclusions that follow from the collective analysis of Figs.(6)--(10) initially may seem counterintuitive.
On one hand maximal attained $E_{\parallel}$ amplitudes drop off as $1/m_i$ (Figs.(6)--(7)). On the other hand, electron flow
maximal attained amplitudes do not depend on $m_i$ (they all are circa $0.03c$, see Fig.(8)), while 
ion flow maximal attained amplitudes   drop off as $1/m_i$ (Figs.~(9)--(10)). Thus one might expect that more massive ions
should produce a bigger $E_{\parallel}$ (since separation of electron and ion fluids is the source
of $E_{\parallel}$ and {\it that} separation is expected to be largest in the case of more massive ions, as they are slower). 
In fact, this is what would be expected if the polarisation drift produced by the driving
IC wave is the cause of parallel electric field generation \cite{glq99,glm04}. The latter two references use the following polarisation drift current:
\begin{equation}
j_{\perp} = \frac{m_i n_i}{B^2} \frac{\partial E_{\perp}}{\partial t},
\end{equation}
where symbols have their usual meaning. The latter equation implies that  $E_{\parallel}$ then should increase with $\propto m_i$.
However, in Figs.(6)--(7) we see completely opposite $E_{\parallel} \propto 1/ m_i$ scaling.
These results can be interpreted (reconciled) as following: (i) ion dynamics plays no role in the $E_{\parallel}$
generation, i.e. polarisation drift has no effect in contrary to the claims of Refs. \cite{glq99,glm04}; 
(ii) decrease in the generated parallel electric field amplitude with the increase of the mass ratio $m_i/m_e$
is caused by the fact that $\omega_d = 0.3 \omega_{ci} \propto 1/m_i$ is decreasing, and hence the electron fluid can
effectively "short-circuit" (recombine with) the slowly 
oscillating ions, hence producing smaller $E_{\parallel}$ which also scales exactly as $1/m_i$.

In summary, indeed, electron and ion fluid separation is causing $E_{\parallel}$ generation, but polarisation drift current
produced by the driving IC wave plays no role.

Interestingly, by comparing Figs.~(8) and (9) we gather that electron fluid is 
efficiently accelerated by the generated $E_{\parallel}$ to the
velocities of up to $0.03c$, while
ion fluid due to its larger inertia is much less mobile ($0.0001c - 0.0007c$). 
This confirms yet another
conclusion that was made in Refs.\cite{tss05a,tss05b} which employed full kinetic simulation.

It should be noted that since here we use two-fluid approach the generated $E_{\parallel}$ cannot
change the distribution function, which obviously remains Maxwellian, while in the previous kinetic simulation of
a similar system it produced bumps in the distribution function as the electrons residing on the magnetic field lines with the density gradients
get efficiently accelerated (see e.g. Fig.(4) in Ref.\cite{tss05b}).

\begin{figure}[]
\resizebox{\hsize}{!}{\includegraphics{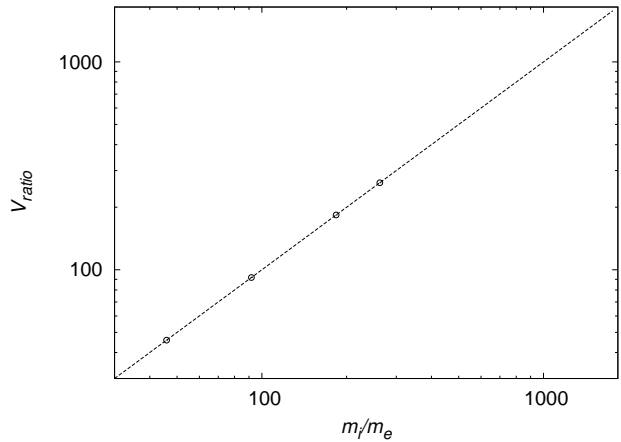}} 
\caption{The ratio of the final attained electron and ion flow amplitudes (within 3 periods of the driving ion cyclotron wave)
as a function of mass ratio, i.e. $V_{ratio}=  \max(|V_{ez}(x,z,t_{final})|) /  \max(|V_{iz}(x,z,t_{final})|)$ vs. $m_i/m_e$. 
The dashed line corresponds to the fit which has a slope of 1 (note the direct correlation). This is a log-log plot.}
\end{figure}

\section{Summary}

We studied the generation of parallel electric fields by
means of propagation of IC waves in the plasma with the
transverse density inhomogeneity. By using simpler, than kinetic \cite{glq99,glm04,tss05a,tss05b}, 
two-fluid, cold plasma linearised equations, we show for the first time that $E_{\parallel}$ generation 
can be understood by an analytic equation that couples $E_{\parallel}$  to the transverse electric field.
It should be noted that the generation of $E_{\parallel}$  is a generic feature of plasmas with the transverse 
density inhomogeneity and in a different context this was known for decades in the
laboratory plasmas \cite{cm93,r82}. We prove that the minimal model required to reproduce the previous kinetic results
of $E_{\parallel}$ generation is the two-fluid, cold plasma approximation in the linear regime. In the latter, the generated 
$E_{\parallel}$  amplitude attains values of  $14$ Vm$^{-1}$ for plausible solar coronal parameters and realistic mass ratio of $m_i/m_e=1836$.
By considering the numerical solutions for $(V_{ez}-V_{iz})$, we have shown that the cause of 
$E_{\parallel}$ appearance is the electron and ion flow separation (which is not the same as electrostatic charge separation)
induced by the transverse density inhomogeneity. 

We also investigate how $E_{\parallel}$ generation is affected by the mass ratio and found
that amplitude attained by $E_{\parallel}$ (within 3 periods of the driving ion cyclotron wave) decreases linearly as 
$\propto 1/m_i$. This result contradicts to the earlier suggestion by G\'enot et al (1999, 2004)
that the cause of $E_{\parallel}$ generation is the polarisation drift of the driving wave, which would suggest $E_{\parallel} \propto m_i$ scaling.
Increase in mass ratio does not affect the final 
parallel (magnetic field aligned) speed attained by electron fluid. However, parallel ion
velocity decreases linearly as  $1/ m_i$, this means that the parallel 
velocity ratio of electrons and ions scales directly as $m_i$. 

 It should be noted that when plasma density is homogeneous, no $E_{\parallel}$ generation takes place; and this is corroborated both by
 numerical simulations (not presented here) and agrees with the Eq.(6) (when $n=const$, the  RHS of Eq.(6) is zero at all times as $E_x$ and $B_y$ do not 
 propagate obliquely).
 Our model also correctly reproduces the previous kinetic results 
\cite{glq99,glm04,tss05a,tss05b} that only electrons are accelerated (along the background magnetic field),
while ions practically show no acceleration.

\section{appendix}

Animations 1, 2, and 3 show movies corresponding to Figs.(3), (4) and (5) respectively.
Note that horizontal axis indicates 500 grids, while the real simulation value is 2500. This is simply to reduce movie size, i.e.
every 5-th point along the field was included in these movies.

\acknowledgements 

The author is supported by Nuffield Foundation (UK) through an award to newly appointed 
lecturers in Science, Engineering and Mathematics (NUF-NAL 04), University of Salford Research Investment 
Fund 2005 grant, and Science and Technology Facilities Council (UK) standard grant. 
The author would like to thank Dr Grigory Vekstein for useful comments at the Manchester-Salford joint seminar. 

{\bf Note added in proof}: The typical values of the Dreicer electric field on the corona is a few 
$\times10^3$ V m$^{-1}$ \cite{t06b}, 
which implies the obtained $E_{\parallel}$ in our model exceeds the Dreicer value 
by at least four orders of magnitude, 
perhaps enabling the electron run away regime. This would imply that our model is more relevant to the 
acceleration of solar wind, rather than solving the coronal heating problem. Essentially acceleration of 
electrons would dominate over the heating as such. However, this seems uncertain because electron and 
ion fluid separation cannot go on (build up) forever, and some sort of discharge should eventually take place. 
At any rate, similar kinetic simulations have shown \cite{glm04} (see their figure 11) thatwave energy is converted 
into particle energy on timescales of $10^3 \omega_{pe}^{-1}$ 
(mind that the latter number is likely to be dependent 
on the mass ratio $m_i/m_e$).

\bibliography{dt_07}

\end{document}